# Application of slingshot model to the giant radio galaxy DA240

(Brief : DA 240 in slingshot model)

*by*

S. Muthumeenal & Dilip G Banhatti[@]

[@][dilip.g.banhatti@gmail.com]

School of Physics, Madurai Kamaraj University, Madurai 625021, India

**Abstract / Summary.** We attempt a slingshot model interpretation of the unusual association of some 1&1/3 dozen nonstellar galaxian objects around the parent optical galaxy of the giant radio galaxy DA 240 (= 0748.6+55.8 (J2000)). Similar interpretation may be possible for another large radio galaxy 3C 31 (= NGC 383 = 0104.6+32.1 (1950.0)).

**Motivation.** Spectroscopic observations of most of the 32 optical objects brighter than about 16 mag within half a degree of DA 240 parent galaxy have been reported by Peng et al (2004). Sixteen objects show nearly the same redshift as the DA 240 parent ($z = 0.0358$). Most of the others have lower redshifts, and are probably outlying members of Zwicky cluster Zw 0756.1+5616. The objects with nearly DA 240 redshift show significant radio emission, are mostly along DA 240 radio major axis, and their redshifts systematically vary along this axis, implying dynamical time scale of 1 Gyr. Interpretation within beam model of active galaxies runs into problems on several counts, while the unusual features of these observations fit quite naturally into slingshot model.

**Beam and slingshot models for active galaxies.** In *beam model* (e.g., Begelman et al 1984 & references therein) the central 'engine' harbours a deep gravitational potential (possibly a supermassive spinning black hole) which generates twin oppositely directed magnetoplasma beams or jets along the black hole accretion disk spin axis, travelling with at least mildly relativistic bulk speed. The observed structures in active galaxies and their nuclei arises from the interaction of these hydromagnetic jets progressively with the surrounding nuclear, interstellar and intergalactic / intracluster media. The highly relativistic magnetoplasma particles emit mainly nonthermal radiation in the entrained magnetic field. This synchrotron radiation is observed in different polarizations and from x-ray to radio part of the electromagnetic radiation spectrum, but mainly in radio. (See e.g., Banhatti 1998 & references therein.) A double radio galaxy emits predominantly radio energy from two elongated regions called radio lobes, one on each of two sides of the optical parent galaxian object. Bulk supersonic speeds of the twin jets through the surrounding medium leads to shocks creating radio hotspots, especially at radio lobe ends.

In *slingshot model* (Saslaw et al 1974) the densely populated supermassive active galactic nucleus ejects two or more massive objects or black holes in two diametrically opposite directions due to gravitational dynamical instability. The black hole trajectories are seen as jets and lobes. The black holes may escape from the supermassive galaxian nucleus or may oscillate relative the nucleus. (See, e.g., Banhatti 1998 & references therein.)

*Basic differences between beam and slingshot models* : (1) In slingshot model the ejected black holes move substantially, while beam model has a single black hole (or possibly a binary) more or less stationary at the nucleus. (2) In the twin beam model the source of energy remains in the active galactic nucleus of the (extended) radio galaxy, the energy being transported continuously to the lobes, which can be very far away from the nucleus, while in slingshot model the sources of energy are massive black hole accretion disk systems which move within the lobes as they locally emit synchrotron energy, so that no long distance energy transport is needed in this 'dressed slingshot' (Lin & Saslaw 1977).

**Slingshot model details and application to some radio galaxies.** This section briefly presents some relevant details of slingshot model and its application to some radio galaxies. The literature used for this survey: Valtonen (1979, 1999), Valtonen & Heggie (1978), Bridle et al (1989), Mikkola & Valtonen (1990), Borcherds & McCauley (1993), Valtonen et al (1994), Valtonen & Heinämäki (2000). In slingshot model supermassive black holes are thrown out from an active galaxian nucleus (AGN) via the gravitational 3-body process. Conserving linear momentum, a single black hole escapes in one direction and a binary in the opposite direction, each producing radio emission along its trajectory. Numerically integrating orbits in the model AG(N) potential and fitting simple functions through the computed points one gets

$(r / r_0) = (2.89 - 2.4 ((t/t_0) - 1)^2)^{1/2} - 0.7, \quad 0 < (t / t_0) < 2$

where r is the distance from the AGN, and

$r_0 = 0.85 \{1 - (V_e/V_{esc})^2\}^{-5/3} (M/(3 \times 10^{12} M_\odot))$ Kpc,

$t_0 = 0.75 (r_0/Kpc)^{1.3} (M/(3 \times 10^{12} M_\odot))^{-0.25}$ Myr.

M is the AGN mass and $M_\odot$ is the solar mass. The escape velocity is

$V_{esc} = 3110 (M/(3 \times 10^{12} M_\odot))^{0.25}$ km/s

These expressions are within 10% of numerically calculated orbits for $0.8 \leq (V_e/V_{esc}) < 1$. Different radio sources modeled in slingshot theory include 3C 465, 3C 83.1B, 3C 288, 3C 277.3, 3C 388, Cen A, 3C 390.3, 3C 405, 3C 219, 3C 129 and 3C 274. The dynamical time scales needed are 10 Myr to 1 Gyr. Schematic orbits are shown as line diagrams. Convolution with appropriate emission function gives radio brightness distribution. (See the references listed earlier.)

**Beam and slingshot models for double radio sources.** In beam theory large scale jets represent particle flow channels, while in slingshot model they are trails left behind by the outgoing black hole and its backward pointing beam. Trails are pressure confined and essentially stationary, while flow channels are seen as dynamic entities.

The frequency of detecting jets and counter-jets in double radio galaxies and quasars goes against the simplest versions of beam model, while it can be readily explained in slingshot model.

Jets which do not originate from the galactic nucleus (e. g., 3C 338) are problematic in beam model but are expected in slingshot model.

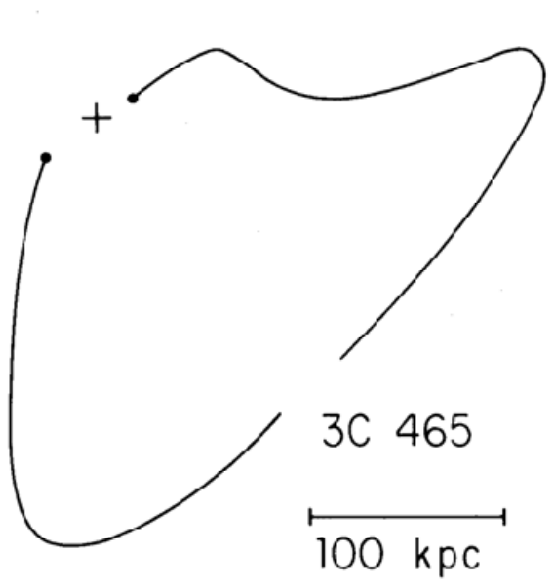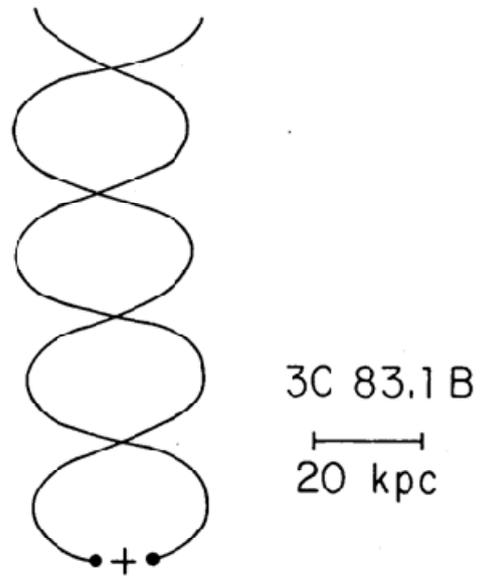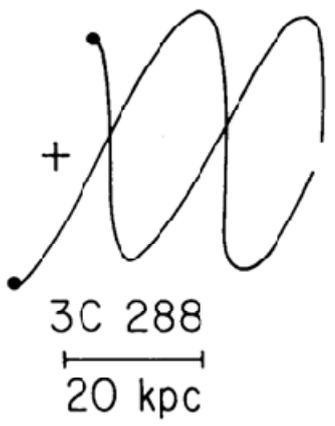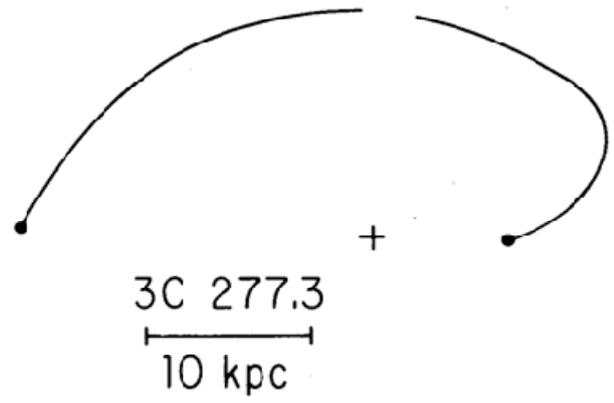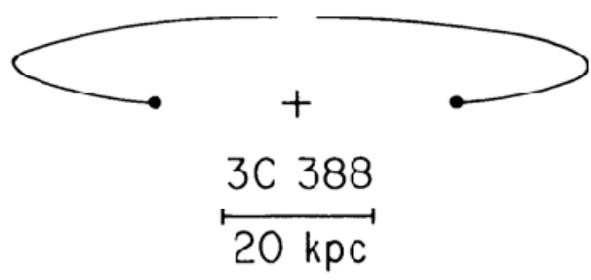

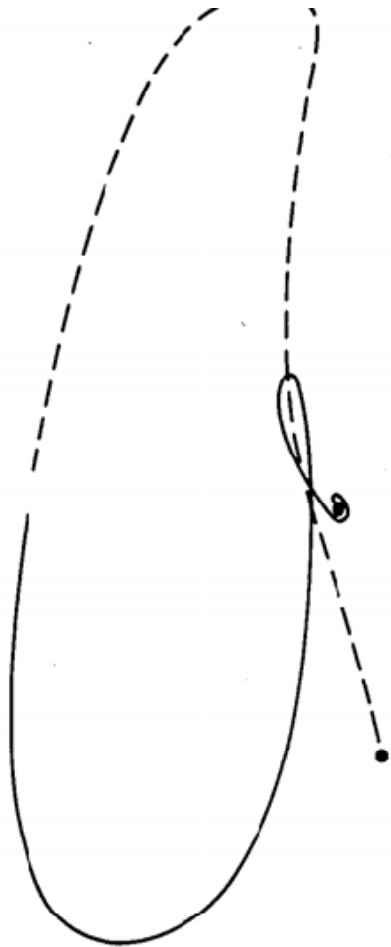

CEN A
200 kpc

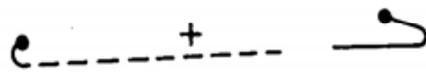

3C 390.3
200 kpc

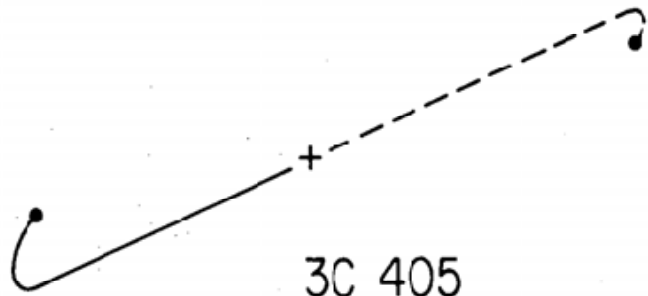

3C 405
100 kpc

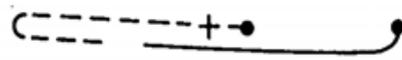

3C 219
300 kpc

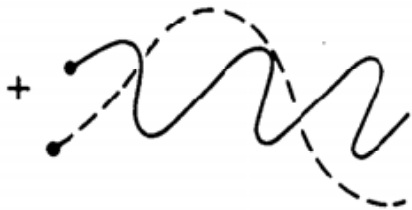

3C 129
50 kpc

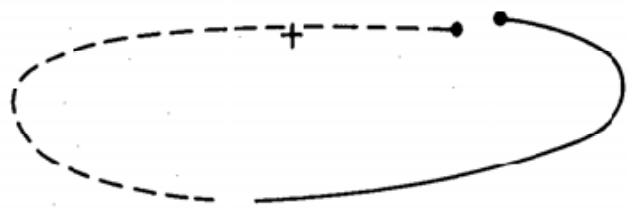

3C 274
2 kpc

Double hotspots in radio lobes sometimes present a problem in beam theory, especially when there are indications of current particle acceleration in both hotspots. In slingshot model double hotspots are a standard feature.

The alignment of hotspots across the nucleus of the galaxy is good in view of the jets and the off-axis positions of the hotspots. This gives a strong indication that hotspots are the primary movers and jets are only secondary phenomena. This is so in slingshot model but not in beam model.

The advantage of slingshot model is that it is an integral part of the general body of astrophysical theory which describes the merger-evolution of galaxies in cosmological setting. Therefore, it is fully calculable and results may be directly compared with observations.

**DA 240 features which call for slingshot model.**

(1)The nonstellar galaxian objects along DA 240 radio major axis must be interpreted in beam model as members of the same group of galaxies as the DA 240 parent. Polarization observations of DA 240 lobes imply number density n ≈ $10^{-5}$ /cc which is too low for galaxy formation. In slingshot model these massive objects have been flung out from the DA 240 AGN, and are at different points of the orbit(s), so can be in any region, of high or low density.

(2)Both DA 240 lobes have roughly symmetrically placed hotspots, but they are very asymmetric in strength, the eastern one about 50 times brighter than its western counterpart. The reason for this brightness contrast is unclear in beam model. In slingshot theory the hotspots are dressed black holes (Lin & Saslaw 1977), symmetrically placed due to linear momentum balance, with their own twin-jet systems that may have very different synchrotron emissivities.

(3)Of the 16 nonstellar galaxian objects with redshifts similar to DA 240 parent, 10 are aligned along the radio lobes while 6 lie along roughly orthogonal axis. In beam theory there is no direct connection between this grouping and the radio galaxy jets or beams. In slingshot theory there is an orbital plane with concentration near the AGN, and due to orbital instability, massive objects are ejected in a narrow cone normal to and on both sides of the central flat nuclear concentration. For such a scenario for DA 240, the 6 objects / galaxies that lie in directions roughly normal to the double radio source axis should be by far the more massive component of the total mass of the DA 240 group. Perhaps this can be verified by future spectroscopic mass estimates.

(4)Monte Carlo simulations show that the chance that 17 objects are aligned as observed is about $3\times10^{-4}$, indicating a causal connection, mysterious in beam model, but quite natural in slingshot mechanism. A stable dynamical axis over time scale much longer than synchrotron emission age is also consistent with slingshot model.

(5)Projecting onto the radio major axis the location of each of 11 objects on or near the axis, and plotting the redshift difference relative to the median as Δv vs the distance D along the axis gives a correlation of slope Δv/D ≈ 425 km/s/Mpc, with correlation coefficient -0.74 (Peng et al 2004). This calls for an explanation, not naturally possible in beam theory, but maybe possible in slingshot model.

*Acknowledgements*. The work reported here formed part of S. Muthumeenal's M.Phil. project in 2007. DGB thanks UGC, New Delhi for financial support.

-x-0-x-